\documentclass[11pt]{llncs}
\usepackage{cite}
\usepackage{amssymb}
\usepackage{amsmath}
\usepackage{graphicx}
\usepackage{fullpage}

\newcommand{\ceil}[1]{\left\lceil{#1}\right\rceil}

\renewcommand{\mod}[0]{\text{ mod }}

\newcommand{\fract}{\mathrm{frac}}

\title{Linear Probing with Constant Independence}

\author{Anna Pagh \and Rasmus Pagh \and Milan Ru\v zi\' c}

\institute{IT University of Copenhagen, Rued Langgaards Vej 7, 2300 K{\o}benhavn S, Denmark.}

\date{\today}

\begin{document}
\maketitle

\begin{abstract}
Hashing with linear probing dates back to the 1950s, and is among the most studied algorithms.
In recent years it has become one of the most important hash table organizations since it uses the cache of modern computers very well.
Unfortunately, previous analysis rely either on complicated and space consuming hash functions, or on
the unrealistic assumption of free access to a truly random hash function. Already Carter and Wegman, in their seminal paper on universal hashing, raised the question of extending their analysis to linear probing.
However, we show in this paper that linear probing using a pairwise independent family may have expected {\em logarithmic\/} cost per operation.
On the positive side, we show that 5-wise independence is enough to ensure constant expected time per operation. This resolves the question of finding a space and time efficient hash function that provably ensures good performance for linear probing.
\end{abstract}


\section{Introduction}

Hashing with linear probing is perhaps the simplest algorithm for storing and accessing a set of keys that obtains nontrivial performance. Given a hash function $h$, a key $x$ is inserted in an array by searching for the first vacant array position in the sequence $h(x), h(x)+1, h(x)+2, \dots$ (Here, addition is modulo $r$, the size of the array.) Retrieval of a key proceeds similarly, until either the key is found, or a vacant position is encountered, in which case the key is not present in the data structure. Linear probing dates back to 1954, but was first analyzed by Knuth in a 1963 memorandum~\cite{oai:CiteSeerPSU:49006} now considered to be the birth of the area of analysis of algorithms~\cite{prodinger-preface}. Knuth's analysis, as well as most of the work that has since gone into understanding the properties of linear probing, is based on the assumption that $h$ is a truly random function. In 1977, Carter and Wegman's notion of universal hashing~\cite{CW} initiated a new era in the design of hashing algorithms, where explicit and efficient ways of choosing hash functions replaced the unrealistic assumption of complete randomness. In their seminal paper, Carter and Wegman state it as an open problem to ``Extend the analysis to [...] double hashing and open addressing.''\footnote{Nowadays the term ``open addressing'' refers to any hashing scheme where the data structure is an array containing only keys and empty locations. However, Knuth used the term to refer to linear probing in~\cite{oai:CiteSeerPSU:49006}, and since it is mentioned here together with the double hashing probe sequence, we believe that it refers to linear probing.}

\subsection{Previous results using limited randomness}

The first analysis of linear probing relying only on limited randomness was given by Siegel in~\cite{STOC::SchmidtS1990,ncstrl.nyu_cs//TR1995-686}. Specifically, he shows that $O(\log n)$-wise independence is sufficient to achieve essentially the same performance as in the fully random case. However, another paper by Siegel~\cite{Siegel2004} shows that evaluation of a hash function from a $O(\log n)$-wise independent family requires time $\Omega(\log n)$ unless the space used to describe the function is $n^{\Omega(1)}$. A family of functions is given that achieves space usage $n^{\epsilon}$ and constant time evaluation of functions, for any $\epsilon > 0$. However, this result is only of theoretical interest since the associated constants are very large (and growing exponentially with~$1/\epsilon$).

A potentially more practical method due to Dietzfelbinger (seemingly described in the literature only as a ``personal communication'' in~\cite{d-cuckoo-jour}) can be used to achieve characteristics similar to those of linear probing, still using space $n^\epsilon$. This method splits the problem into many subproblems of roughly the same size, and simulates full randomness on each part. Thus, the resulting solution would be a {\em collection\/} of linear probing hash tables.

A significant drawback of both methods above, besides a large number of instructions for function evaluation, is the use of random accesses to the hash function description. The strength of linear probing
is that for many practical parameters, almost all lookups will incur only a single cache miss. Performing
random accesses while computing the hash function value may destroy this advantage.

\subsection{Our results}

We show in this paper that linear probing using a pairwise independent family may have expected {\em logarithmic\/} cost per operation. Specifically, we resolve the open problem of Carter and Wegman by showing that linear probing insertion of $n$ keys in a table of size $2n$ using a function of the form $x\mapsto ((ax+b) \mod p) \mod 2n$, where $p=4n+1$ is prime and we randomly choose $a\in [p]\backslash \{0\}$ and $b\in [p]$, requires $\Omega(n\log n)$ insertion steps in expectation for a worst case insertion sequence (chosen independently of $a$ and $b$). Since the total insertion cost equals the total cost of looking up all keys, the expected average time to 
look up a key in the resulting hash table is $\Omega(\log n)$. The main observation behind the proof is that
if $a$ is the multiplicative inverse (modulo $p$) of a small integer $m$, then inserting a certain set that consists of two intervals has expected cost $O(n^2/m)$.

On the positive side, we show that 5-wise independence is enough to ensure constant amortized expected time per insertion, for load factor $\alpha= n/r$ bounded away from~1. This implies that the expected average time for a successful search is constant. We also show a constant expected time bound for unsuccessful searches.
Our proof is based on a new way of bounding the cost of
linear probing insertions, in terms of ``fully loaded'' intervals $I$, where the number of probe sequences starting in $I$ is at least $|I|$.

Our analysis of linear probing gives a bound of $n(1+O(\frac{\alpha}{(1-\alpha)^2}))$ steps for $n$ insertions. This implies a bound of $1+O(\frac{\alpha}{(1-\alpha)^2})$ steps on average for successful searches. 
For higher values of $\alpha$, these bounds are a factor $\Omega(\frac{1}{1-\alpha})$ higher than for linear probing with full independence.
To get a better dependence on $\alpha$ we introduce a class of open addressing methods
called {\em blocked probing}, which includes a special kind of bidirectional linear probing. 
For this scheme, which shares the cache friendliness of traditional linear probing, we get the same dependence on $\alpha$ (up to constant factors) as for full independence, 
again using only 5-wise independent hash functions.
In addition, for blocked linear probing we bound the expected cost of any single insertion, deletion, or
unsuccessful search, rather than the average cost of a sequence of operations.
For successful searches, we analyze the average cost.
In this case, we have a bound for 4-wise independent families as well.

\subsection{Significance}

Several recent experimental studies~\cite{cuckoo-jour,HeilemanLuo2005,conf/wae/BlackMQ98} have found linear probing to be clearly the fastest hash table organization for moderate load factors (30-70\%). While linear probing operations are known to require more instructions than those of other open addressing methods, the fact that they access an interval of array entries means that linear probing works very well with modern architectures for which sequential access is much faster than random access (assuming that the elements we are accessing are each significantly smaller than a cache line, or a disk block, etc.). However, the hash functions used to implement linear probing in practice are heuristics, and there is no known theoretical guarantee on their performance. Since linear probing is particularly sensitive to a bad choice of hash function, Heileman and Luo~\cite{HeilemanLuo2005} advice {\em against\/} linear probing for general-purpose use. Our results imply that simple and efficient hash functions, whose description can be stored in CPU registers, can be used to give 
provably good performance.


\section{Preliminaries}

\subsection{Notation and definitions}

Define $[x] = \{0,1,\ldots, x-1\}$.
Throughout this paper $S$ denotes a subset of some universe $U$,
and $h$ will denote a function from $U$ to $R=[r]$. 
We denote the elements of $S$ by $\{x_1, x_2,\ldots, x_n\}$, and refer to the elements of $S$ as {\em keys}. We let $n=|S|$, and $\alpha=n/r$.
For any integers $x$ and $a$ define $x \ominus a = x - (x \mod a)$.
The function $x \mapsto x - \lfloor x \rfloor$ is denoted by $\fract(x)$

A family $\mathcal{H}$ of functions from $U$ to $R$ is $k$-wise independent if for any $k$ distinct elements $x_1,\dots,x_k\in U$ and $h$ chosen uniformly at random from $\mathcal{H}$, the random variables $h(x_1),\dots,h(x_k)$ are independent. We say that $\mathcal{H}$ is $\epsilon$-approximately uniform if function values for random $h\in\mathcal{H}$ have $L_\infty$ distance at most $\epsilon$ from the uniform distribution, i.e., for any $x\in U$ and any $y\in [r]$ it holds that $|\Pr\{h(x)=y\}-1/r| \leq \epsilon$.
We note that in some papers, the notion of $k$-wise independence is stronger in that it is required that function values are uniform on $[r]$. However, many interesting $k$-wise independent families have a slightly nonuniform distribution, and we will provide analysis for such families as well.

Let $Q$ be a subset of the range $R$.
By $Q+a$ we denote the translated set $\{(a+y)\mod r\ |\ y\in Q\}$.
We will later use sets of form $Q+h(x)$, for a fixed $x$ and $Q$ being an interval.
An {\em interval\/} (modulo~$r$) is a set of the form $[b]+a$, for integers $a$ and $b$.

Here we introduce a function which simplifies statements of some upper bounds.
Define function $T(\alpha, \epsilon)$ with domain $\alpha \in [0,1)$, 
$\epsilon \in [0, \frac{1-\alpha}{\alpha})$, and the value given  by
\begin{displaymath}
T(\alpha, \epsilon) = 
 \min\left\{\frac{{5.2\alpha(1+\epsilon)^2}}{(1-(1+\epsilon)\alpha)^2} + \frac{4}{9\alpha} - 1,\
  \frac{{3\alpha^2(1+\epsilon)^2}}{(1-(1+\epsilon)\alpha)^4}\Big(2 + \frac{4}{9\alpha}\Big)\right\} \enspace .
\end{displaymath}
Remark that $T(\alpha, \epsilon) = O\big(\frac{\alpha(1+\epsilon)^2}{(1-\alpha)^2}\big)$.


\subsection{Hash function families}

Alon et al.~\cite{MR88a:68043} observed that the family of degree $k-1$ polynomials in any finite field is $k$-wise independent. Specifically, for any prime $p$ we may use the field defined by arithmetic modulo~$p$ to get a family of functions from $[p]$ to $[p]$ where a function can be evaluated in time $O(k)$ on a RAM, assuming that addition and multiplication modulo $p$ can be performed in constant time. 
To obtain a smaller range $R=[r]$ we may map integers in $[p]$ down to $R$ by a modulo $r$ operation.
This of course preserves independence, but the family is now only close to uniform. Specifically, it has $L_\infty$ distance less than $1/p$ from the uniform distribution on $[r]$.

A recently proposed $k$-wise independent family of Thorup and Zhang~\cite{DBLP:conf/soda/ThorupZ04}
has uniformly distributed function values in $[r]$. From a theoretical perspective (ignoring constant factors) 
it is inferior to Siegel's highly independent family~\cite{Siegel2004}, since the evaluation time depends on $k$. We mention it here because it is the first construction that makes $k$-wise independence truly competitive with popular heuristics, for small $k>2$, in terms of evaluation time. The construction for 4-wise independence is particularly efficient in practice. Though this is not stated in~\cite{DBLP:conf/soda/ThorupZ04}, it is not hard to verify that the same construction in fact gives 5-wise independence, and thus our analysis will apply.


\subsection{Probability bounds for fully loaded intervals}

Here we state a lemma that is essential for our upper bound results, described in 
Section~\ref{sect -- linear probing upper bound} and Section~\ref{sect - blocked probing}.
The technical proof is deferred to the final section.

\begin{lemma} \label{lemma - overloaded}
Let $\mathcal{H}$ be a $4$-wise independent and $\frac{\epsilon}{r}$-approximately uniform 
family of functions which map $U$ to $R$,
with $\epsilon < 1 - \frac{2}{n}$.
If $h$ is chosen uniformly at random from $\mathcal{H}$,
then for any $Q\subset R$ of size $q$,
\begin{displaymath}
\Pr\{|h(S)\cap Q| \geq \alpha q(1+\epsilon) + d)\} \leq 
     \frac{3\alpha^2q^2 + \alpha q}{d^4} (1+\epsilon)^2
\end{displaymath}
If the family of functions is $5$-wise independent and $\epsilon < 
     \min\left\{ 1 - \frac{2}{n},\frac{1-\alpha}{\alpha}\right\},$
then for any fixed $x\in U\setminus S$,
\begin{displaymath}
\Pr\{ |h(S)\cap (Q+h(x))| \geq q)\} \leq 
  \left(3\alpha^2q^{-2} + \alpha q^{-3}\right) \frac{(1+\epsilon)^2}{\big(1-(1+\epsilon)\alpha\big)^4} \enspace .
\end{displaymath}
\end{lemma}


\section{Lower bound for pairwise independence}

Consider the following family of functions, introduced by Carter and Wegman~\cite{CW} as a first example of a universal family of hash functions: 
$$\mathcal{H}(p,r) = \{ x\mapsto ((ax+b) \mod p) \mod r \; | \; 0<a<p,\, 0\leq b < p\}$$
where $p$ is any prime number and $r\leq p$ is any integer.
Functions in $\mathcal{H}(p,r)$ map integers of $[p]$ to $[r]$. 
We slightly modify $\mathcal{H}(p,r)$ to be pairwise independent and have uniformly distributed 
function values. Let $\hat{p}=\ceil{p/r}r$, and define a function $g$ as follows: $g(y,\hat{y})=\hat{y}$ if $\hat{y}\geq p$, and $g(y,\hat{y})=y$ otherwise. For a vector $v$ let $v_i$ denote the $i+1$st component (indexes starting with zero). We define:
$$\mathcal{H}^*(p,r) = \{ x\mapsto g((ax+b) \mod p, v_x) \mod r \; | \; 0\leq a<p,\, 0\leq b < p,\, v\in [\hat{p}]^p\}$$
Our lower bound on the performance of linear probing will apply to both $\mathcal{H}$ and $\mathcal{H}^*$. This gives an example of a very commonly used hash function family that does {\em not\/} yield expected constant time per operation for linear probing for a fixed, worst case set. It also shows that pairwise independence is not a sufficient condition for a family to work well with linear probing, and thus complements our upper bounds for 5-wise (and higher) independence.

\begin{lemma}[Pairwise independence]\label{lem:pairwise}
For any pair of distinct values $x_1,x_2\in [p]$, and any $y_1,y_2\in [r]$, if $h$ is chosen uniformly at random from $\mathcal{H}^*(p,r)$, then
$$\Pr\{h(x_1)=y_1 \wedge h(x_2)=y_2\} = 1/r^2 \enspace . $$
\end{lemma}
\begin{proof}
We will show something stronger than claimed, namely that the family 
$$\mathcal{H}^{**}=\{ x\mapsto g((ax+b) \mod p, v_x) \; | \; 0\leq a<p,\, 0\leq b < p,\, v\in [\hat{p}]^p\}$$
is pairwise independent and has function values uniformly distributed in $[\hat{p}]$. Since $r$ divides $\hat{p}$ this will imply the lemma. Pick any pair of distinct values $x_1,x_2\in [p]$, and consider a random function $h\in \mathcal{H}^{**}$. Clearly, $v_{x_1}$ and $v_{x_2}$ are uniform in $[\hat{p}]$ and independent. Also, it follows by standard arguments~\cite{CW} that
$(a x_1+b) \mod p$ and $(a x_2+b) \mod p$ are uniform in $[p]$ and independent. We can think of the definition of $h(x)$ as follows: The value is $v_x$ unless $v_x\in [p]$, in which case we substitute $v_x$ for another random value in $[p]$, namely $(a x+b) \mod p$. It follows that hash function values are uniformly distributed, and pairwise independent. \qed
\end{proof}

To lower bound the cost of linear probing we use the following lemma:
\begin{lemma}\label{lem:intersection_costbound}
Suppose that $n$ keys are inserted in a linear probing hash table of size $r$ with probe sequences starting at $i_1,\dots,i_n$, respectively. Further, suppose that $I_1,\dots,I_\ell$ is any set of intervals (modulo~$r$) such that we have the multiset equality $\cup_j \{i_j\} = \cup_j I_j$. Then the total number of steps to perform the insertions is at least $$\sum_{1\leq j_1 < j_2 \leq \ell} |I_{j_1}\cap I_{j_2}|^2/2 \enspace .$$
\end{lemma}
\begin{proof}
We proceed by induction on $\ell$. Since the number of insertion steps in independent of the order of insertions, we may assume that the insertions corresponding to $I_\ell$ occur last.
By the induction hypothesis, the total number of steps to do all preceding insertions is at least 
$\sum_{1\leq j_1 < j_2 \leq \ell-1} |I_{j_1}\cap I_{j_2}|^2/2$. Let $S_j$ denote the set of keys corresponding to $I_j$. For any $j<\ell$, and any $x\in S_\ell$ with probe sequence starting in $I_j\cap I_\ell$, the insertion of $x$ will pass all keys of $S_j$ with probe sequences starting in $I_j\cap I_\ell$. This means that at least $|I_{j}\cap I_{\ell}|^2/2$ steps are used during the insertion of the keys of $S_\ell$ to pass locations occupied by keys of $S_j$. Summing over all $j<\ell$ and adding 
to the bound for the preceding insertions finishes the induction step. \qed
\end{proof}

\begin{theorem}\label{thm:lower}
For $r=\ceil{p/2}$ there exists a set $S\subseteq [p]$, $|S|\leq r/2$, 
such that the expected cost of inserting the elements of $S$ in a linear probing hash table of size $r$ using a hash function chosen uniformly at random from $\mathcal{H}(p,r)$ is 
$\Omega(r\log r)$.
\end{theorem}
\begin{proof}
We first define $S$ as a random variable, and show that when choosing $h$ at random from $\mathcal{H}(p,r)$ the expected total insertion cost for the keys of $S$ is $\Omega(r\log r)$. This implies the existence of a fixed set $S$ with at least the same expectation for random $h\in \mathcal{H}(p,r)$. Specifically, we subdivide $[p]$ into 8 intervals $U_1,\dots,U_8$, such that $\cup_i U_i = [p]$ and $r/4\geq |U_i| \geq r/4-1$ for $i=1,\dots,8$, and let $S$ be the union of two of the sets $U_1,\dots,U_8$ chosen at random (without replacement). Note that $|S|\leq r/2$, as required.

Consider a particular function $h\in \mathcal{H}(p,r)$ and the associated values of $a$ and $b$.
Let $\hat{h}(x)=(ax+b)\mod p$, and let $m$ denote the unique integer in $[p]$ such that $am \mod p = 1$ 
(i.e., $m=a^{-1}$ in GF($p$)). Since $\hat{h}$ is a permutation on $[p]$, the sets $\hat{h}(U_i)$, $i=1,\dots,8$, are disjoint.
We note that for any $x$, $\hat{h}(x+m) = (\hat{h}(x)+1)\mod p$. 
Thus, for any $k$, $\hat{h}(\{x,x+m,x+2m,\dots,x+km\})$ is an interval (modulo $p$) of length $k+1$. 
This implies that for all $i$ there exists a set $\hat{L}_i$ of $m$ disjoint intervals such that $\hat{h}(U_i) = \cup_{I\in L_i} I$. 
Similarly, for all $i$ there exists a set $L_i$ of at most $m+1$ intervals (not necessarily disjoint) such that we have the multiset equality $h(U_i) = \cup_{I\in L_i} I$. 
Since all intervals in $\cup_i \hat{L}_i$ are disjoint, an interval in $\cup_i L_i$ can intersect at most 
two other intervals in $\cup_i L_i$. We now consider two cases: 

1. Suppose there is some $i$ such that $\sum_{I_1,I_2\in L_i, I_1\ne I_2} |I_1\cap I_2|\geq r/16$.
Then with constant probability $U_i\subseteq S$, and we apply the bound of Lemma~\ref{lem:intersection_costbound}. 
The sum is minimized if all $O(m)$ nonzero intersections have the same size, $\Omega(r/m)$. Thus Lemma~\ref{lem:intersection_costbound} implies that the number of insertion steps is $\Omega(r^2/m)$.

2. Now suppose that for all $i$, $\sum_{I_1,I_2\in L_i, I_1\ne I_2} |I_1\cap I_2| < r/16$. Note that any value in $[r-1]$ is contained in exactly two intervals of $\cup_i L_i$, and by the assumption at most half occur in two intervals of $L_i$ for some $i$. Thus there exist $i_1,i_2$, $i_1\ne i_2$, such that $|h(U_{i_1})\cap h(U_{i_2})|=\Omega(r)$. With constant probability we have $S=U_{i_1} \cup U_{i_2}$. We now apply Lemma~\ref{lem:intersection_costbound}. Consider just the terms in the sum of the form $|I_1\cap I_2|^2/2$, where $I_1\in L_{i_1}$ and $I_2\in L_{i_2}$. As before, this sum is minimized if all $O(m)$ intersections have the same size, $\Omega(r/m)$, and we derive an $\Omega(r^2/m)$ lower bound on the number of insertion steps.

For a random $h\in \mathcal{H}(p,r)$, $m$ is uniformly distributed in $\{1,\dots,p\}$ (the map $a\mapsto a^{-1} $ is a permutation of $\{1,\dots,p\}$). Therefore, the expected total insertion cost is $\Omega(\frac{1}{p} \sum_{m=1}^p r^2/m) = \Omega(r^2\log p/p) = \Omega(r\log r) \enspace .$
\qed
\end{proof}

\begin{corollary}
Theorem~\ref{thm:lower} holds also if we replace $\mathcal{H}(p,r)$ by $\mathcal{H}^*(p,r)$. In particular, pairwise independence is not a sufficient condition for linear probing to have expected constant cost per operation.
\end{corollary}
\begin{proof}
Consider the parameters $a$, $b$, and $v$ of a random function in $\mathcal{H}^*(p,r)$.
Since $r=\ceil{p/2}$ we have $\hat{p}=p+1$, and $(p/\hat{p})^p > 1/4$. Therefore, with constant probability it holds that $a\ne 0$ and $v\in [p]^p$.
Restricted to functions satisfying this, the family $\mathcal{H}^*(p,r)$ is identical to $\mathcal{H}(p,r)$. Thus, the lower bound carries over (with a smaller constant). By Lemma~\ref{lem:pairwise}, $\mathcal{H}^*$ is pairwise independent with uniformly distributed function values. \qed
\end{proof}

We remark that the lower bound is tight. 
A corresponding $O(n\log n)$ upper bound can be shown by applying the analysis of Section~\ref{sect -- linear probing upper bound} and using Chebychev's inequality to bound the probability of a fully loaded interval,
rather than using the 4th moment inequality as in Lemma~\ref{lemma - overloaded}.


\section{Linear probing with 5-wise independence} \label{sect -- linear probing upper bound}

We analyze the cost of performing $n$ insertions into an empty table of size $r = \frac{n}{\alpha}$.
From this, a bound on the cost of a successful search (of a random element in the set) can be derived.
The cost of insertions and the average search cost do not depend on the order of insertions --
or equivalently, on the policy of placing elements being inserted.
We assume that the following policy is in effect: if $x$ is the new element to be inserted,
place $x$ into the first slot $h(x)+i$ that is either empty or 
contains an element $x'$ such that $h(x') \notin h(x)+[i+1]$.
If $x$ is placed into a slot previously occupied by $x'$ then the probe sequence continues as if $x'$
is being inserted.
The entire procedure terminates when an empty slot is found.

\begin{theorem} \label{theorem - insertions for lin prob}
Let $\mathcal{H}$ be a $5$-wise independent and $\frac{\epsilon}{r}$-approximately uniform 
family of functions which maps $U$ to $R$,
with $\epsilon < \min\left\{1 - \frac{2}{n}, \frac{1-\alpha}{\alpha}\right\}$.
When linear probing is used with a hash function chosen uniformly at random from $\mathcal{H}$, 
the expected total number of probes made by a sequence of 
$n$ insertions into an empty table is less than $n(1 + T(\alpha, \epsilon))$.
\end{theorem}
\begin{proof}
For every $x_i\in S$, let $d_i$ be the displacement of $x$, i.e. the number such that $x_i$ resides 
in slot $(h(x_i) + d_i) \mod r$.
The entire cost of all insertions is equal to $\sum_{i=1}^n (1 + d_i)$.
From the way elements are inserted, we conclude that, for $1\leq i\leq n$ and $1 \leq l \leq d_i$, 
every interval $h(x_i)+[l]$ is \emph{fully loaded}, meaning that at least $l$ elements of 
$S \setminus \{x_i\}$ hash into it.
Let $A_{il}$ be the event that the interval of slots $h(x_i)+[l]$ is fully loaded.
Then,
\begin{displaymath}
E(d_i) = \sum_{k=1}^r \Pr\left\{ d_i\geq k \right\} \leq \sum_{k=1}^r \Pr\left(\bigcap_{l=1}^{k} A_{il}\right) \leq \sum_{j=0}^{\lfloor\lg r\rfloor} 2^{j} \cdot \Pr(A_{i\, 2^j})
\end{displaymath}
Lemma \ref{lemma - overloaded} gives us an upper bound on $\Pr(A_{i\, 2^j})$.
However, for small lengths and not small $\alpha$ the bound is useless, so then we will simply use the trivial upper bound of~1.
Let $K = \frac{{3\alpha^2(1+\epsilon)^2}}{(1-(1+\epsilon)\alpha)^4}$.
We first consider the case $K\geq 1$.
Denoting $j_* = \big\lceil \frac{1}{2}\lg K\big\rceil$ we have
\begin{eqnarray*}
E(d_i) &\leq& 2^{j_*} - 1 + \sum_{j=j_*}^{\lg r} 2^{j}\left(\frac{K}{2^{2j}} + \frac{K}{3\alpha \cdot 2^{3j}} \right)
             = 2^{j_*}-1 + K \sum_{j=j_*}^{\lg r} 2^{-j} + \frac{K}{3\alpha} \sum_{j=j_*}^{\lg r} 2^{-2j} \\
     &<& 2^{j_*}-1 + \frac{K}{2^{j_*}}\frac{1}{1-\frac{1}{2}} + 
                   \frac{K}{3\alpha \cdot 4^{j_*}}\frac{1}{1-\frac{1}{4}} \\
     &\leq&\sqrt{K}\cdot 2^{1-\fract(\lg \sqrt{K})} - 1 + 2\sqrt{K} \cdot 2^{-(1-\fract(\lg \sqrt{K}))} +
              \frac{4}{9\alpha}\enspace .
\end{eqnarray*}
The last expression is not larger than $3\sqrt{K} + \frac{4}{9\alpha} - 1$, 
because $2^t + 2\cdot 2^{-t} \leq  3$, for $t \in [0,1]$.
Doing an easier calculation without splitting of the sum at index $j_*$ 
gives $E(d_i) < K(2 + \frac{4}{9\alpha})$.
The bound $3\sqrt{K} + \frac{4}{9\alpha} - 1$ is higher than the bound $K(2 + \frac{4}{9\alpha})$
when $K < 1$, and thus we can write $E(d_i) < T(\alpha, \epsilon)$.
\qed
\end{proof}


\section{Blocked probing} \label{sect - blocked probing}

In this section we propose and analyze a family of open addressing methods, 
containing among other a variant of bidirectional linear probing.
Suppose that keys are hashed into a table of size $r$ by a function $h$.
For simplicity we assume that $r$ is a power of two.
Let $V_j^i=\{j, j+1, \ldots, j + 2^i-1\}$ where $j$ is assumed to be a multiple of $2^i$.
Intervals $V_j^i$ may be thought of as sets of references to slots in the hash table.
In a search for key $x$ intervals $V_j^i$ that enclose $h(x)$ are examined in the order of increasing $i$.
More precisely, $V_{h(x)}^0$ is examined first; 
if the search did not finish after traversing $V_{h(x) \ominus 2^i}^i$,
then the search proceeds in the untraversed half of $V_{h(x) \ominus 2^{i+1}}^{i+1}$.
The search stops after traversal of an interval if either of the following three cases hold:
\begin{itemize}
\item[a)] key $x$ was found,
\item[b)] the interval contained empty slot(s),
\item[c)] the interval contained key(s) whose hash value does not belong to the interval.
\end{itemize}
In case (a) the search may obviously stop immediately on discovery of $x$ --
there is no need to traverse through the rest of the interval.

Traversal of unexamined halves of intervals $V_j^i$ may take different concrete forms -- 
the only requirement is that every slot is probed exactly once.
From a practical point of view, a good choice is to probe slots sequentially 
in a way that makes the scheme a variant of bidirectional linear probing.
This concrete scheme defines a probe sequence that in probe numbers $2^i$ to $2^{i+1}-1$ inspects
either slots 
\begin{displaymath}
(h(x) \ominus 2^i + 2^i,\ h(x) \ominus 2^i + 2^i + 1, \ldots,\ h(x) \ominus 2^i + 2^{i+1} - 1)
\end{displaymath}
\begin{displaymath}
\mathrm{or}\quad (h(x) \ominus 2^i - 1,\ h(x) \ominus 2^i - 2, \ldots,\ h(x) \ominus 2^i - 2^{i})
\end{displaymath}
 depending on whether $h(x) \mod 2^i = h(x) \mod 2^{i+1}$ or not.
A different probe sequence that falls in this class of methods, but is not sequential, is 
$(x,j) \mapsto h(x)\ \mathrm{xor}\ j$, with $j$ starting from 0.

\paragraph{Insertions.} Until key $x$ which is being inserted is placed in a slot, 
the same probe sequence is followed as in a search for $x$.
However, $x$ may be placed in a non-empty slot if its hash value is closer to the slot number 
in a special metric which we will now define 
(it is not hard to guess what kind of metric should that be for the above given search procedure to work).
Let $d(y_1, y_2) = \min \{i\ |\ y_2 \in V_{y_1 \ominus 2^i}^i\}$.
The value of $d(y_1, y_2)$ is equal to the position of the most significant bit in which $y_1$ and $y_2$ differ.
If during insertion of $x$ we encounter a slot $y$ containing key $x'$ then 
key $x$ is put into slot $y$ if $d(h(x), y) < d(h(x'), y)$.
In an implementation there is no need to evaluate $d(h(x), y)$ values every time.
We can keep track of what interval $V_{h(x) \ominus 2^i}^i$ is being traversed at the moment 
and check whether $h(x')$ belongs to that interval.

When $x$ is placed in slot $y$ which was previously occupied by $x'$, a new slot for $x'$ has to be found.
Let $i=d(h(x'), y)$.
The procedure now continues as if $x'$ is being inserted and we are starting with traversal of 
$V_{h(x') \ominus 2^i}^i \setminus V_{h(x') \ominus 2^{i-1}}^{i-1}$.
If the variant of bidirectional linear probing is used, the traversal may start from position $y$,
which may matter in practice.

\paragraph{Deletions.}
After removal of a key we have to check if the new empty slot can be used to bring some keys 
closer to their hash values, in terms of metric $d$.
If there is an additional structure among stored elements, like in the bidirectional linear probing variant,
some elements may be repositioned even though the corresponding values of metric $d$ do not decrease.
Let $x$ be the removed key, $y$ be the slot in which it resided, and $i=d(h(x), y)$.
There is no need to examine $V_{h(x) \ominus 2^{i-1}}^{i-1}$.
If $V_{h(x) \ominus 2^i}^i \setminus V_{h(x) \ominus 2^{i-1}}^{i-1}$ contains another empty slot then 
the procedure does not continue in wider intervals.
If it continues and an element gets repositioned then 
the procedure is recursively applied starting from the new empty slot.

It is easy to formally check that appropriate invariants hold and 
that the above described set of procedures works correctly.

\subsection{Analysis}

We analyze the performance of operations on a hash table of size $r$ when this class of 
probe sequences is used.
Suppose that the hash table stores an arbitrary fixed set of $n=\alpha r$ elements.
Let $C_\alpha^U$, $C_\alpha^I$, $C_\alpha^D$, and $C_\alpha^S$ be the random variables that respectively represent: 
the number of probes made during an unsuccessful search for a fixed key,
the number of probes made during an insertion of a fixed key,
the number of probes made during a deletion of a fixed key,
and the number of probes made during a successful search for a random element from the set.
In the symbols for the random variables we did not explicitly include marks for the fixed set and 
fixed elements which are used in the operations, but they have to be implied.
The upper bounds on the expectations of $C_\alpha^\Xi$ variables, which are given by the following theorem,
do not depend on choices of those elements.

\begin{theorem} \label{theorem - expect for fixed elem}
Let $\mathcal{H}$ be a $5$-wise independent and $\frac{\epsilon}{r}$-approximately uniform 
family of functions which map $U$ to $R$,
with $\epsilon < \min\left\{1 - \frac{2}{n}, \frac{1-\alpha}{\alpha}\right\}$.
For a load factor $\alpha<1$, blocked probing with a hash function chosen 
uniformly at random from $\mathcal{H}$ provides the following expectations: 
$E(C_\alpha^U) < 1 + T(\alpha, \epsilon)$,
$E(C_\alpha^I) < 1 + 2T(\alpha, \epsilon)$,
$E(C_\alpha^D) < 1 + 2T(\alpha, \epsilon)$, and
\begin{displaymath}
E(C_\alpha^S) < \left\{ \begin{array}{ll}
   1 + \left(\alpha^2 + \frac{\alpha}{3}\right)\frac{4(1+\epsilon)^2}{(1-(1+\epsilon)\alpha)^3} \ ,&
        \mathrm{if\ } \alpha \leq \frac{0.3}{1+\epsilon} \\
  \frac{0.915}{\alpha(1+\epsilon)}  + \frac{10.4(1+\epsilon)}{1-(1+\epsilon)\alpha} + \frac{0.673}{\alpha} 
       - 1 + \frac{1}{\alpha}\ln \left((1-(1+\epsilon)\alpha)^{10.4} (\alpha(1+\epsilon))^{8/9} \right)\ , &
    \mathrm{if\ } \alpha > \frac{0.3}{1+\epsilon}
    \end{array} \right.  \enspace . 
\end{displaymath}
\end{theorem}
\begin{proof}
Denote by $x$ the fixed element from $U\setminus S$ that 
is being searched/inserted.
Let $\bar{C}_\alpha^U$ be the random variable that takes value $2^i$ when $2^{i-1}< C_\alpha^U \leq 2^i$,
$0\leq i \leq \lg r$.
We can write $\bar{C}_\alpha^U = 1 + \sum_{i=1}^{\lg r} 2^{i-1} T_i$, where $T_i$ is an indicator variable whose
value is $1$ when at least $2^{i-1}+1$ probes are made during the search.
Let $A_j$ be the event that the interval of slots 
$V_{h(x) \ominus 2^j}^j$ is \emph{fully loaded}, meaning that at least $2^j$ elements are hashed into the interval.
Then $T_i=1$ when the chosen function $h$ is in $\bigcap_{j=0}^{i-1} A_j$.
We get an overestimate of $E(\bar{C}_\alpha^U)$ with 
$E(\bar{C}_\alpha^U) \leq 1 + \sum_{i=1}^{\lg r} 2^{i-1} \Pr(A_{i-1})$.
The sum $\sum_{i=0}^{\lg r} 2^{i} \Pr(A_{i})$ appeared in the proof of 
Theorem \ref{theorem - insertions for lin prob},
so we reuse the upper bound found there.

We now move on to analyzing insertions.
Let $\bar{C}_\alpha^I$ be the random variable that takes value $2^i$ when $2^{i-1}< C_\alpha^I \leq 2^i$,
$0\leq i \leq \lg r$.
Variable $C_\alpha^U$ gives us the slot where $x$ is placed, but we have to consider
possible movements of other elements.
If $x$ is placed into a slot previously occupied by key $x'$ from a ``neighboring''
interval $V_{h(x) \ominus 2^{i+1}}^{i+1} \setminus V_{h(x) \ominus 2^{i}}^{i}$,
then as many as $2^{i}$ probes may be necessary to find a place for $x'$ in $V_{h(x) \ominus 2^{i}}^{i}$,
if there is one.
If entire $V_{h(x) \ominus 2^{i+1}}^{i+1}$ is fully loaded, then as many as $2^{i+1}$ additional probes may 
be needed to find a place within $V_{h(x) \ominus 2^{i+2}}^{i+2} \setminus V_{h(x) \ominus 2^{i+1}}^{i+1}$,
and so on.
In general -- and taking into account all repositioned elements -- 
we use the following accounting to get an overestimate of $E(\bar{C}_\alpha^I)$:
for every fully loaded interval $V_{h(x) \ominus 2^{i}}^{i}$ we charge $2^i$ probes,
and for every fully loaded neighboring interval $V_{h(x) \ominus 2^{i+1}}^{i+1} \setminus V_{h(x) \ominus 2^{i}}^{i}$ 
we also charge $2^i$ probes.
The probability of a neighboring interval of length $2^i$ being full is equal to $\Pr(A_i)$.
As a result, 
$E(\bar{C}_\alpha^I) \leq 1 + \sum_{i=0}^{\lg r - 1} 2^i \cdot 2\Pr(A_i) < 1 + 2T(\alpha, \epsilon)$.

Reasoning for deletions is similar.

A bound on $E(C_\alpha^S)$ can be derived from the bound on $E(C_\alpha^I)$ by using the observation
$E(C_\alpha^S) \leq \sum_{i=0}^{n-1} E(C_{i/r}^I)$, as in linear probing
\footnote{The equality in the relation need not hold in general for blocked probing. 
It holds in the bidirectional linear probing variant.}.
The solution to the equation $\frac{{3z^2(1+\epsilon)^2}}{(1-(1+\epsilon)z)^4}=1$, over $z\in[0,1]$, is
$\approx \frac{0.29}{1+\epsilon}$.
When substituting $E(C_{i/r}^I)$ with $1 + 2T(\frac{i}{r}, \epsilon)$, we choose to use 
the second function from the expression for $T(\alpha, \epsilon)$ when $i < \frac{0.3}{1+\epsilon}r$,
and to use the first function for higher $i$ 
(recall the discussion from the proof of Theorem \ref{theorem - insertions for lin prob}).
The obtained sums are suitable for approximation by integrals.
After some technical work we end up with the claimed bound on $E(C_\alpha^S)$.
\qed
\end{proof}

For higher values of $\alpha$, the dominant term in the upper bounds is $O((1-\alpha)^{-2})$ 
(except for successful searches).
The constants factors in front of term $(1-\alpha)^{-2}$ are relatively high compared to
standard linear probing with fully random hash functions.
This is in part due to approximative nature of the proof of Theorem \ref{theorem - expect for fixed elem},
and in part due to tail bounds that we use, which are weaker than those for fully independent families.
In the fully independent case, the probability that an interval of length $q$ is fully loaded is less than
$e^{q(1-\alpha+\ln\alpha)}$, according to Chernoff-Hoeffding bounds \cite{Chernoff52, Hoeffding63}.
Plugging this bound into the proof of Theorem \ref{theorem - expect for fixed elem} would give, e.g.,
\begin{equation} \label{eq - bound on C_alpha^U for full indep}
E(C_\alpha^U) < 1 + \frac{e^{1-\alpha+\ln\alpha}}{\ln 2 \cdot |1-\alpha + \ln\alpha|} \enspace .
\end{equation}
For $\alpha$ close to $1$, a good upper bound on (\ref{eq - bound on C_alpha^U for full indep})
is $1 + \frac{2}{\ln 2}(1-\alpha)^{-2}$.
The constant factor here is $\approx 2.88$, as opposed to $\approx 5.2$ from the statement of 
Theorem \ref{theorem - expect for fixed elem}.
As $\alpha$ gets smaller, the bound in (\ref{eq - bound on C_alpha^U for full indep}) gets further
below $1 + \frac{2}{\ln 2}(1-\alpha)^{-2}$.

As we will show in the next theorem, 4-wise independence is sufficient to get good performance of
successful searches.
We will more directly bound the sum of displacements of all elements of $S$.
However, if we tried to be very precise in calculating an upper bound,
calculations would become complex and seemingly impossible to keep at the level of elementary functions
(e.g. summing $\sum_i \frac{2^{2i}}{((1-\alpha)2^i + 1)^3}$.
Instead, we do a relatively simple calculation.
A bound with concrete constants is given only for the case $\alpha \geq 0.8$,
as an illustration.

\begin{theorem}
Let $\mathcal{H}$ be a $4$-wise independent and $\frac{\epsilon}{r}$-approximately uniform 
family of functions which map $U$ to $R$,
with $\epsilon < \min\left\{1 - \frac{2}{n}, \frac{1-\alpha}{\alpha}\right\}$.
When blocked probing is used with a hash function chosen uniformly at random from $\mathcal{H}$, 
then $E(C_\alpha^S) < 1 + O(\frac{1}{1-\alpha})$.
If $\alpha \geq 0.8$, an upper bound on $E(C_\alpha^S)$ is 
$\frac{6(1+\epsilon)}{1-(1+\epsilon)\alpha} - 2.7$.
\end{theorem}
\begin{proof}
It is sufficient to bound the expectation of $\sum_{i=1}^n(1 +  \sum_{l=0}^{\lg r - 1} 2^{l}T_{il})$,
where $T_{il}$ is an indicator variable whose value is 1 if element $x_i$ is not placed
inside interval $V_{h(x_i) \ominus 2^{l}}^{l}$.
We will estimate $E(\sum_{i=1}^n T_{il})$.
For easier exposition, we introduce the symbol $\bar{\alpha} = (1+\epsilon)\alpha$.

Let $w$ be a variable with domain $[(1-\bar{\alpha})2^l + 1, n]$, 
let $z$ be a variable with domain $(1,+\infty)$, and set $K_l = (3\bar{\alpha}^2 2^{2l} + \bar{\alpha} 2^l)$.
For every interval $V_j^l$ we may bound the number of elements that have overflowed as follows:
make a fixed charge of $w - (1-\bar{\alpha})2^l - 1$ elements plus the expected additional overflow given by:
\begin{displaymath}
\sum_{j=0}^{r/2^l - 1} \sum_{\lambda=0}^{\lg \lceil\frac{n}{w}\rceil} w(z^{\lambda+1} - z^\lambda) 
                     \cdot \Pr\{|h(S)\cap V_{2^l j}^l| \geq  \bar{\alpha}2^l + z^\lambda w\}
\end{displaymath}
\begin{displaymath}
\quad  <  \sum_{j=0}^{r/2^l - 1} \sum_{\lambda = 0}^{\infty} 
              w z^\lambda(z-1) \frac{K_l}{(z^{\lambda}w)^4} \leq \frac{K_l}{w^3}\frac{r}{2^l} \frac{z-1}{1-z^{-3}}
\end{displaymath}
The above inequality holds for any $z > 1$.
Therefore, from $\lim_{z\to 1} \frac{z-1}{1-z^{-3}} = \frac{1}{3}$ it follows
$E(\sum_{i=1}^n T_{il}) < \frac{r}{2^l}(w - (1-\bar{\alpha})2^l - 1 + \frac{K_l}{3w^3})$.
The minimum of the upper bound is reached for $w=\max\{\sqrt[4]{K_l}, (1-\bar{\alpha})2^l + 1\}$.

We will focus on the case $\bar{\alpha} \geq 0.8$.
One of the simplifications that we make is to use $w = \sqrt[4]{K_l}$ for 
$0\leq l < l_* = \big\lfloor \lg\frac{\sqrt{3}\bar{\alpha}}{(1-\bar{\alpha})^2} \big\rfloor$.
It can be shown that $\sqrt[4]{K_l} - (1-\bar{\alpha})2^l - 1 > 0$, for $0\leq l < l_*$.
Unless $\bar{\alpha}$ is very close to 1, index $l_*$ is at most one less than 
the optimal splitting index in this case (often, it is optimal).
Then,
\begin{eqnarray*}
E\left(\sum_{l=0}^{l_*-1}2^l \sum_{i=1}^n T_{il}\right) &<& 
 r \sum_{l=0}^{l_*-1}\left(\frac{4}{3}\sqrt[4]{K_l} - (1-\bar{\alpha})2^l - 1\right) \\
 &\leq& r\sum_{l=0}^{l_*-1} \frac{4}{3}\sqrt{\bar{\alpha} 2^l}\sqrt[4]{3 + \frac{1}{\bar{\alpha} 2^l}} 
        - r(1-\bar{\alpha})(2^{l_*}-1) - r\cdot l_*\\
 &\leq& r\frac{4 \sqrt{\bar{\alpha}} }{3}\frac{2^{l_*/2}}{\sqrt{2}-1}\sqrt[4]{4.25} - 
        r(1-\bar{\alpha})2^{l_*} - 3.8r\cdot \quad (\mathrm{because\ } \bar{\alpha} \geq 0.8) \ .
\end{eqnarray*}
For the second part of the sum, we have
\begin{eqnarray*}
E\left(\sum_{l=l_*}^{\lg r - 1}2^l \sum_{i=1}^n T_{il}\right) &<& 
     r \sum_{l=l_*}^{\lg r - 1}\frac{K_l}{3(1-\bar{\alpha})^3 2^{3l}}\\
 &\leq& r \sum_{l=l_*}^{\lg r - 1} \left(\frac{\bar{\alpha}^2}{(1-\bar{\alpha})^3 2^{l}} + 
                                   \frac{\bar{\alpha}}{3(1-\bar{\alpha})^3 2^{2l}}\right) \\
 & < &  \frac{r}{(1-\bar{\alpha})^3} \left(\frac{2\bar{\alpha}^2}{2^{l_*}} + 
      \frac{4\bar{\alpha}}{9\cdot 2^{2l_*}}\right) \enspace .
\end{eqnarray*}
Merging the sums and substituting $2^{l_*}$ results in
\begin{displaymath}
E\left(\sum_{l=0}^{\lg r - 1}2^l \sum_{i=1}^n T_{il}\right)
 < \frac{\bar{\alpha} r}{1-\bar{\alpha}}(6.083\cdot 2^{-t/2} - 1.731\cdot 2^{-t} + 1.155\cdot 2^{t}) 
    - 3.8r + (1-\bar{\alpha})r\frac{0.149}{\bar{\alpha}} \ ,
\end{displaymath}
where $t = \fract\big(\lg \frac{\sqrt{3}\bar{\alpha}}{(1-\bar{\alpha})^2}\big)$.
It can be shown that $6.083\cdot 2^{-t/2} - 1.731\cdot 2^{-t} + 1.155\cdot 2^{t} < 6$.
Substituting $\frac{0.149(1-\bar{\alpha})}{\bar{\alpha}^2}(1+\epsilon)$
with the maximum value over $\bar{\alpha}\in[0.8,1]$ yields the upper bound for this case.

Performance for other values of $\alpha$ can be analyzed in a similar simple way, 
by suitably choosing indexes $l$ for which $w$ is set to $\sqrt[4]{K_l}$ (e.g. sometimes it will be 
for $l\in \{1,2,3\}$)
When $\alpha$ is very close to 0, analysis for $\alpha=c$, with $c$ being a small constant,
is applied.
\qed
\end{proof}

Achieving a significantly smaller constant factor is planned for future work.
A more elegant proof is required.

\section{The proof of Lemma \ref{lemma - overloaded}}

Let $X_i$ be the indicator random variable that has value 1 
iff $h(x_i)\in Q$, $1\leq i \leq n$.
By our assumptions on $\mathcal{H}$,
the variables $X_i$ are $4$-wise independent and 
$\frac{q}{r}(1-\epsilon) \leq\Pr\{X_i = 1\} \leq \frac{q}{r} (1 + \epsilon)$.
We use the 4th moment inequality:
\begin{displaymath}
\Pr\{|X - \mu| \geq d\} \leq \frac{E((X-\mu)^4)}{d^4} \ ,
\end{displaymath}
with  $X = \sum_{i=1}^n X_i$, $\mu=E(X)$.
In terms of raw moments, the 4th central moment is expressed as
\begin{displaymath}
E((X-\mu)^4) = E(X^4) - 4\mu E(X^3) + 6\mu^2 E(X^2) - 3\mu^4 \ .
\end{displaymath}
We will expand the raw moments and express them in a form that will
allow later cancellation of high-order terms.
The simplest is the second moment:
\begin{eqnarray*}
E(X^2) &=& E\Big(\Big(\sum_{i=1}^n X_i\Big)^2\Big) = E\Big(\sum_i X_i^2 + \sum_{i\neq j} X_iX_j\Big)
           = E \Big(\sum_i X_i + \sum_{i\neq j} X_iX_j\Big) \\
 &=& E(X) + \sum_{i=1}^n\sum_{j\neq i} E(X_i X_j) = E(X) + \sum_{i=1}^n E(X_i)\sum_{j\neq i} E(X_j) \\
 &=& E(X) + \sum_{i=1}^n E(X_i)(E(X) - E(X_i)) = E(X) + E(X)^2 - \sum_i (E(X_i))^2 \enspace .
\end{eqnarray*}
The equality $X_i^2 = X_i$ is true because $X_i$ is an indicator variable.
Also, $E(X_iX_j) = E(X_i)E(X_j)$, $i\neq j$, because $X_i$ and $X_j$ are independent.
Defining $\sigma_k = \sum_i (E(X_i))^k$, the identity is written more succinctly as:
$E(X^2) = \mu + \mu^2 - \sigma_2$.

Define the predicate $\Delta$ such that $\Delta(a_1,a_2,\ldots,a_k)$ is true iff $a_1,\ldots,a_k$ are all distinct.
For the third moment we have, using independence of any three indicator variables:
\begin{eqnarray*}
E(X^3) &=& E\Big(\Big(\sum_{i=1}^n X_i\Big)^3\Big) 
             = E\Big(\sum_i X_i^3 + \sum_{i\neq j} 3X_i^2 X_j + \sum_{\Delta(i,j,k)} X_iX_jX_k\Big) \\
 &=& E \Big(\sum_i X_i + \sum_{i\neq j} 3X_iX_j + \sum_{\Delta(i,j,k)} X_iX_jX_k\Big) \\
 &=& \mu + 3\mu^2 - 3\sigma_2 + \sum_{i=1}^n E(X_i)\sum_{j\neq i} E(X_j) \sum_{\Delta(i,j,k)} E(X_k)\\
 &=& \mu + 3\mu^2 - 3\sigma_2 + \sum_{i} E(X_i)\sum_{j\neq i} E(X_j) (E(X) - E(X_j) - E(X_i))\\
 &=& \mu + 3\mu^2 - 3\sigma_2 + \sum_{i} E(X_i) \Big((\mu - E(X_i))^2 - \sum_{j\neq i} E(X_j)^2\Big)\\
 &=& \mu + 3\mu^2 - 3\sigma_2 + \mu^3 -3\mu\sigma_2 + 2 \sigma_3 \enspace .
\end{eqnarray*}
Expanding $E(X^4)$ in the same way, using 4-wise independence (and reusing some of the previous calculations) yields:
\begin{eqnarray*}
E\Big(\Big(\sum_{i=1}^n X_i\Big)^4 \Big) 
 &=& E\Big(\sum_i X_i^4 + \sum_{i\neq j} 4X_i^3 X_j + \sum_{i\neq j} 3X_i^2 X_j^2 
    + \sum_{\Delta(i,j,k)} 6X_i^2X_jX_k + \sum_{\Delta(i,j,k,l)} X_iX_jX_kX_l \Big) \\
 &=&  \mu + 7\mu^2 - 7\sigma_2 + 6\mu^3 - 18\mu\sigma_2 + 12\sigma_3
    + \mu^4 - 6\mu^2\sigma_2 + 8\mu\sigma_3 + 3(\sigma_2)^2 - 6\sigma_4 \enspace .
\end{eqnarray*}
Combining all the identities results in:
\begin{displaymath}
E((X-\mu)^4) = \mu + 3\mu^2 - 7\sigma_2 - 6\mu\sigma_2 + 12\sigma_3
     + 3(\sigma_2)^2 - 6\sigma_4 \enspace .
\end{displaymath}
Because of $E(X_i) \leq 1$, it follows that $\mu \geq \sigma_2 \geq \sigma_3$.
Thus $-3\mu\sigma_2 + 3(\sigma_2)^2 \leq 0$, and therefore
\begin{displaymath}
E((X-\mu)^4) < \mu + 3\mu^2 - 3\mu\sigma_2 + 12\sigma_3 
 \leq \mu + 3\mu^2 - \sigma_2(3\alpha q(1-\epsilon) - \frac{q}{r}(1+\epsilon)) < \mu + 3\mu^2  \enspace.
\end{displaymath}
The last inequality is true due to the condition that $\epsilon < 1 - \frac{2}{n}$,
from the statement of the lemma.
Noticing that $3\mu^2 + \mu < 3\alpha^2q^2(1+\epsilon)^2 + \alpha q(1+\epsilon)$,
finishes the first part of the lemma.

For the second part, observe that in the subfamily $\{h\in \mathcal{H}\ |\ h(x)=a\}$,
for a fixed $a$, is 4-wise independent and $\frac{\epsilon}{r}$-approximately uniform.
Using the result of the first part by conditioning on the event $h(x)=a$, 
and then applying the total probability theorem yields the claimed inequality.

\small
\bibliographystyle{abbrv}
\bibliography{linear}

\end{document}